\documentclass[12pt]{article}

\usepackage{mathtools}
\usepackage{tikz-cd}
\usepackage{adjustbox}
\usepackage{capt-of}
\usepackage[utf8]{inputenc}
\usepackage{textcomp}
\usepackage{amsmath}
\usepackage[shortlabels]{enumitem}
\usepackage{chetnew}
\usepackage{tikz}
\usepackage{amssymb,amsfonts,mathrsfs,dsfont,yfonts,bbm}\usetikzlibrary{calc}
\usepackage{xcolor}
\usepackage{braket}
\usepackage[normalem]{ulem}
\usepackage{import}
\usepackage{cancel}
\usepackage{booktabs}
\usepackage{varioref}
\usepackage{verbatim}
\usepackage{hyperref}
\usepackage{graphicx}
\usepackage{bm}
\usepackage{tikz-cd}

\newcommand{\cI}{\mathcal{I}}
\newcommand{\cE}{\mathcal{E}}
\newcommand{\cC}{\mathcal{C}}

\newcommand{\beqn}{\begin{eqnarray}}
\newcommand{\eeqn}{\end{eqnarray}}

\newcommand{\be}{\begin{equation}}
\newcommand{\ee}{\end{equation}}
\newcommand{\bea}{\begin{eqnarray}}
\newcommand{\eea}{\end{eqnarray}}

\newcommand{\CA}{\mathcal{A}}

\newcommand{\CH}{\mathcal{H}}

\newcommand{\CD}{\mathcal{D}}
\newcommand{\CE}{\mathcal{E}}

\newcommand{\CB}{\mathcal{B}}
\newcommand{\CC}{\mathcal{C}}

\newcommand{\CO}{\mathcal{O}}
\newcommand{\CT}{\mathcal{T}}
\newcommand{\CI}{\mathcal{I}}
\newcommand{\CN}{\mathcal{N}}
\newcommand{\CS}{\mathcal{S}}

\newcommand*{\boxcoloro}{orange}
\makeatletter
\newcommand{\boxedo}[1]{\textcolor{\boxcoloro}{%
\tikz[baseline={([yshift=-1ex]current bounding box.center)}] \node [rectangle, minimum width=1ex,rounded corners,draw] {\normalcolor\m@th$\displaystyle#1$};}}
 \makeatother
\newcommand*{\boxcolorr}{red}
\makeatletter
\newcommand{\boxedr}[1]{\textcolor{\boxcolorr}{%
\tikz[baseline={([yshift=-1ex]current bounding box.center)}] \node [rectangle, minimum width=1ex,rounded corners,draw] {\normalcolor\m@th$\displaystyle#1$};}}
 \makeatother
\newcommand*{\boxcolorb}{blue}
\makeatletter
\newcommand{\boxedb}[1]{\textcolor{\boxcolorb}{%
\tikz[baseline={([yshift=-1ex]current bounding box.center)}] \node [rectangle, minimum width=1ex,rounded corners,draw] {\normalcolor\m@th$\displaystyle#1$};}}
 \makeatother
\newcommand*{\boxcolorg}{green}
\makeatletter
\newcommand{\boxedg}[1]{\textcolor{\boxcolorg}{%
\tikz[baseline={([yshift=-1ex]current bounding box.center)}] \node [rectangle, minimum width=1ex,rounded corners,draw] {\normalcolor\m@th$\displaystyle#1$};}}
 \makeatother
 \newcommand*{\boxcolorp}{purple}
\makeatletter
\newcommand{\boxedp}[1]{\textcolor{\boxcolorp}{%
\tikz[baseline={([yshift=-1ex]current bounding box.center)}] \node [rectangle, minimum width=1ex,rounded corners,draw] {\normalcolor\m@th$\displaystyle#1$};}}
 \makeatother
  \newcommand*{\boxcolorc}{cyan}
\makeatletter
\newcommand{\boxedc}[1]{\textcolor{\boxcolorc}{%
\tikz[baseline={([yshift=-1ex]current bounding box.center)}] \node [rectangle, minimum width=1ex,rounded corners,draw] {\normalcolor\m@th$\displaystyle#1$};}}
 \makeatother
  \newcommand*{\boxcolory}{yellow}
\makeatletter
\newcommand{\boxedy}[1]{\textcolor{\boxcolory}{%
\tikz[baseline={([yshift=-1ex]current bounding box.center)}] \node [rectangle, minimum width=1ex,rounded corners,draw] {\normalcolor\m@th$\displaystyle#1$};}}
 \makeatother

\usepackage{amsmath}	

\begin{document}
\title{From Free Fields to Interacting SCFTs \\[3mm] via Representation Theory}

\author{Matthew Buican$^{\diamondsuit}$ and Hongliang Jiang$^{\clubsuit}$}

\affiliation{\smallskip CTP and Department of Physics and Astronomy\\
Queen Mary University of London, London E1 4NS, UK\emails{$^{\diamondsuit}$m.buican@qmul.ac.uk, $^{\clubsuit}$h.jiang@qmul.ac.uk}}

\abstract{We ask when it is possible to construct arbitrary unitary multiplets of the superconformal algebra with eight Poincar\'e supercharges that are compatible with locality from (continuous deformations of) representations in free field theories. We answer this question in two, three, and five dimensions. In four dimensions, we resort to an intricate but self-consistent web of conjectures. If correct, these conjectures imply various new non-perturbative constraints on short multiplets in any local unitary 4d $\CN=2$ superconformal field theory and on an unusual set of related vertex algebras. Throughout, we connect our results with properties of deformations in the space of theories.}

\medskip

\date{August 2023}
\setcounter{tocdepth}{2}
\maketitle

\tableofcontents

\newsec{Introduction}
One of the deepest problems in theoretical physics is to characterize the topology of the space of quantum field theories (QFTs), $\CT_{\rm QFT}$.\footnote{See \cite{Vafa:1988ue,Douglas:2010ic,Gukov:2015qea,Gukov:2016tnp,Gaiotto:2019asa,Zeev:2022cnv} for various potential structures on this space.} In this spirit, it is interesting to understand if all QFTs are connected via continuous deformations to free theories.\footnote{See \cite{Douglas:2010ic} for some interesting ideas in two spacetime dimensions.} A somewhat simpler problem along these same lines is to understand when the irreducible representations, $\mathcal{R}_{\CA}$, of some symmetry algebra, $\CA$, can be realized by free fields. In situations where some of these irreps, $\mathcal{R}'_{\CA}\subseteq \mathcal{R}_{\CA}$, cannot be constructed in free QFTs, it is natural to ask whether they can be continuously deformed to representations realized by free fields. To make these questions somewhat more constrained, we can specialize to local unitary QFTs and choose $\CA$ judiciously.

The main goal of this note is to take steps toward solving this latter question for the case that $\CA$ is the superconformal algebra (SCA) with eight Poincar\'e supercharges in two, three, four, and five spacetime dimensions. As we will see, this question is particularly difficult (and therefore interesting) in four spacetime dimensions. Indeed, we solve this problem (up to some caveats we will describe) in two, three, and five dimensions, but we resort to generating an intricate web of mutually consistent conjectures in four dimensions.\footnote{As we will briefly review, the case in 5d is already solved, via an explicit counterexample, in \cite{Buican:2016hpb}. We include 5d in our discussion here so as to emphasize a pattern that emerges across dimensions.}

These latter conjectures, if correct, give rise to non-trivial constraints on the irreps of $\mathfrak{su}(2,2|2)$ in local unitary 4d $\CN=2$ SCFTs and on their interactions with other symmetries. They also provide some new non-perturbative diagnostics that can feed into the bootstrap program.

Heuristically, we expect the following picture to hold. In lower dimensions, free fields and Lagrangians are powerful: relative to higher dimensions, there are more relevant deformations and therefore a more complicated space of infrared (IR) theories given a particular ultraviolet (UV) theory. Our rough intuition is that, given two unitary representations, $\mathcal{R}_1$ and $\mathcal{R}_2$, of the SCA related by a continuous deformation\footnote{In other words, through deforming some abelian quantum number that is not subject to a quantization condition in the space of theories with eight Poincar\'e supercharges.} and compatible with locality, there is at least one pair of local unitary SCFTs with eight supercharges, $\CT_{1,2}$, realizing these representations and at least one deformation in the space of local unitary QFTs with eight Poincar\'e supercharges that connects $\CT_1\ni \mathcal{R}_1$ and $\CT_2\ni \mathcal{R}_2$.\footnote{Note that this deformation need not involve a standard UV to IR RG flow; for example, we will discuss precisely such a case in 2d where the deformation is along a conformal manifold. We can, among other deformations, also flow \lq\lq up" and \lq\lq down" RG trajectories as in \cite{Gaiotto:2019asa}, back and forth across dimensions, gauge symmetries, and/or tensor in / \lq\lq stack" additional QFTs.} This intuition arises from thinking about superconformal indices, and it also gives a physical meaning to deformations of SCA irreps.\footnote{In fact a stronger conjecture might be true: given any such $\mathcal{R}_{1,2}$ related by a continuous deformation, there is always a local unitary deformation connecting $\CT_{1,2}\ni\mathcal{R}_{1,2}$.} Indeed, this idea puts deformations of SCA irreps on a similar physical footing with the irreps themselves.

In $d=2$ spacetime dimensions, free fields are especially powerful. In this case, not only is it often possible to construct free field realizations of various symmetry representations, but it is also often possible to construct interacting QFT operator algebras (i.e., OPEs) in terms of free fields (e.g., see \cite{Feigin:1981st,Dotsenko:1984nm} and the recent discussion in \cite{Kapec:2020xaj}). Therefore, we expect that in 2d, where, for the purposes of this paper, we specialize to $(4,4)$ SCFTs with the left and right \lq\lq small" $\CN=4$ superconformal algebra (SCA), free fields will be enough to reproduce all representations (short/protected ones as well as long/unprotected ones). Indeed, we will see this is the case.

In $d=3$ spacetime dimensions, the relevant supersymmetry algebra with eight Poincar\'e supercharges is $\CN=4$. As is well known, the corresponding Lagrangians are considerably more powerful than in $4d$ $\CN=2$. For example, many Argyres-Douglas theories in 4d have simple 3d $\CN=4$ Lagrangian descriptions upon reduction on a circle \cite{Xie:2012hs,Buican:2015hsa,Benvenuti:2018bav,Shan:2023xtw}. Moreover, 3d mirror symmetry relates the Higgs branch and Coulomb branch and ensures that free fields have equal power in both sectors (free fields are particularly versatile in the case of Higgs branches due to various non-renormalization theorems). Therefore, it is reasonable to believe that all protected representations of the 3d $\CN=4$ SCA can be constructed via free fields. We will indeed see this statement is correct. However, long multiplets cannot generally be realized in this way. On the other hand, long multiplets can be constructed via continuous deformations of SCA irreps built from free fields. We conjecture this statement is related to the existence of corresponding deformations in the space of 3d $\CN=4$ theories.

In $d=4$ spacetime dimensions, the relevant supersymmetry algebra is $\CN=2$. Here locality plays a particularly interesting role. Indeed, superconformal Ward identities and the averaged null energy condition (ANEC) constrain the spectrum of SCA representations \cite{Manenti:2019kbl,Manenti:2019jds} (see also the non-supersymmetric discussion in \cite{Cordova:2017dhq} and the discussion in \cite{Buican:2014qla}).\footnote{These constraints don't say anything about interacting non-local 4d $\CN=2$ SCFTs. However, we expect the space of such theories to be somewhat more constrained than their equally supersymmetric relatives in three and five dimensions. For example, due to Nahm's classification \cite{Nahm:1977tg}, there are no SCFTs that can support non-local 4d $\CN=2$ SCFTs as half-BPS boundary conditions.} However, the results in \cite{Manenti:2019kbl,Manenti:2019jds} on their own are not enough to show that all short irreps of the 4d $\CN=2$ SCA can be realized by free fields up to continuous deformations. Instead, we will formulate a series of conjectures implying that all irreps (short and long) of the 4d $\CN=2$ SCA can be related to those realized by free fields via continuous deformations. For example, we conjecture the following bound for all local unitary SCFTs\footnote{There is also a conjugate bound for conjugate multiplets with $j\leftrightarrow\bar j$.}
\be
\bar \CC_{R,r(j,\bar j)}: \qquad j\le \bar j+R+1~,
\ee
where these multiplets are the \lq\lq least protected" short irreps of $\mathfrak{su}(2,2|2)$ (we have used the nomenclature of \cite{Dolan:2002zh}). Here $R$ is the $SU(2)_R$ spin, $j$ and $\bar j$ are the left and right Lorentz spins, and $r$ is the $U(1)_r$ quantum number. We also conjecture the following bound for the $r=j-\bar j$ limiting case 
\be
 \hat \CC_{R (j,\bar j)}: \qquad |r|=|j-\bar j|\le R+1~.
\ee
Taking some of the spins to be formally negative, we get $\bar\CB$ and $\bar\CD$ multiplets and recover known bounds on their quantum numbers arising from locality.

While we do not prove the above bounds, we give various mutually compatible and reinforcing pieces of evidence for why we believe they hold. As a byproduct, our conjectures lead to interesting constraints on 4d $\CN=2$ SCFTs and also on a vertex algebra (VA) we call the \lq\lq$\sqrt{\rm VOA}$" VA associated with the vertex operator algebras (VOAs) arising via the 4d/2d correspondence in \cite{Beem:2013sza}.

Finally, we conclude in 5d, with a brief discussion of the $\CN=1$ SCA (we leave an analysis of 6d $(1,0)$ to future work). There we review the results of \cite{Buican:2016hpb} which show that there are short representations of the 5d $\CN=1$ SCA that cannot be realized by free fields but do exist in local interacting theories. In so doing, we complete our picture showing that the power of free fields diminishes as we go up in dimension from $d=2$ to $d=5$.

The plan of the paper is as follows. In the next section we discuss low dimensions (i.e., two and three spacetime dimensions). Then, we move to four dimensions before briefly discussing five dimensions. We conclude with some avenues for future work.

\newsec{Low dimensions}
Before discussing the 4d $\CN=2$ SCA, we study the 2d $(4,4)$ and 3d $\CN=4$ SCAs. The results in these dimensions are stronger than those in higher dimensions (in line with the expectations outlined in the introduction). For example, constraints from locality do not play a significant a role in our discussion below.

Let us first consider 2d $(4,4)$ theories. We will restrict ourselves to study theories with left and right copies of the \lq\lq small" $\CN=4$ SCA. To that end, recall that this SCA has representations of the following type (e.g., see \cite{Lee:2019uen}): $[0]_0$, $[1/2]_{1/2}$, $[R\ge1]_{R}$, and $[R]_{h>R}$. 

\bigskip
\noindent
{\bf Note:} Our notation here differs slightly from that in \cite{Lee:2019uen}: $[R]_h$ denotes an irrep with a primary having $SU(2)_R$ spin $R$ (as opposed to the Dynkin label in \cite{Lee:2019uen}), and $h$ is the (left) conformal scaling dimension. Here we only need to focus on the left-moving sector, as the right-moving sector is subject to similar logic.

\bigskip
\noindent
The first three representations above are short, and the final representation is long. We will give a constructive proof showing all can be realized by free fields.

To understand this statement, consider the $(4,4)$ sigma model with target space a square $T^4$ with radii $r_i$ ($i=1,\cdots,4$). This is an SCFT defined in terms of four pairs
of left- and right-moving bosonic $U(1) $ currents, $j^i(z)$ and $\tilde j^i(\bar z )$  $(i = 1, . . . , 4)$, four pairs of left- and right-moving free real fermions, $\psi^i (z)$ and $\tilde \psi^i (\bar z )$, and exponential (primary) fields, $V_k (z,\bar z )$, labelled by
vectors $k   \in \Gamma$ in  an even self-dual lattice of signature $(4, 4)$ (e.g., see \cite{Volpato:2014zla}).

The four left-moving fermions, $\psi^i$, have $h=1/2$. They are primaries of type $[1/2]_{1/2}$. By considering a product of $n$ such theories, we can clearly construct short representations of type $[n/2]_{n/2}$ for any $n\in\mathbb{N}$. On the other hand, using vertex operators, $V_k$, for the bosons of the $T^4$ theory and moving along the conformal manifold (i.e., tuning the radii), we can construct long multiplets with $R=0$ and $h$ as close to $0$ as desired when going to the decompactified limit of large $r_i$. We can then construct any multiple of this scaling dimension through products of $T^4$ theories at these radii. Repeating the same procedure with arbitrarily many fermions from arbitrarily many decoupled $T^4$ sigma models, we can construct long multiplets with any $SU(2)_R$ spin and dimension. As a result, we can use free fields to construct all allowed small 2d $\CN=4$ SCA representations. We therefore see that

\bigskip
\noindent
{\bf Theorem 1:} Any representation of the \lq\lq small" 2d $(4,4)$ algebra (i.e., both left and right algebras are small $\CN=4$ algebras) can be constructed from free fields.

\bigskip
Next let us consider the 3d $\CN=4$ SCA. Note that theories with this SCA can often be related to 4d $\CN=2$ SCFTs via dimensional reduction on a circle. Intuitively, we expect to be able to construct all short 3d $\CN=4$ SCA representations from free fields.

One naive reason to believe this statement is true is that 3d $\CN=4$ Lagrangians are powerful.\footnote{As discussed in the introduction, this phenomenon is more generally true the lower in dimension one goes.} For example, as touched on in the introduction, (four-dimensional) Argyres-Douglas theories do not have Lagrangians with manifest 4d $\CN=2$ supersymmetry (they can sometimes be related to UV $\CN=2$ Lagrangians via subtle IR scaling limits on the Coulomb branches of Lagrangian theories), but many have dimensional reductions described by straightforward IR limits of 3d $\CN=4$ gauge theories \cite{Xie:2012hs,Buican:2015hsa,Benvenuti:2018bav,Shan:2023xtw}.

However, the above argument is too quick: the UV limits of 3d $\CN=4$ gauge theories are scale invariant but not conformal due to their vector multiplets.\footnote{This statement is already true in the simple case of free Maxwell theory. Indeed, the would-be current for special conformal transformations is not conserved \cite{Jackiw:2011vz}. Moreover, the field strength, $F_{\mu\nu}$, is not a conformal primary \cite{El-Showk:2011xbs}.} Therefore, the short distance theory will have protected multiplets that are not part of the SCA.

Following the discussion in \cite{El-Showk:2011xbs}, one might hope to embed the UV limit of such gauge theories into some larger SCFTs. However, a more direct connection between free vectors and a conformal theory is through RG flow. Indeed, we can take the free vector and couple it to a charged hypermultiplet to produce $\CN=4$ SQED with $N_f=1$. Due to the presence of monopole operators, the IR limit of this theory is the free twisted hypermultiplet \cite{Gaiotto:2008sa}. The existence of such a representation is required by mirror symmetry (more generally, we expect free twisted hypermultiplets at generic points on the Coulomb branch).

In the language of the 3d $\CN=4$ SCA, we see that free hypers and free vectors are connected to the following representations in the nomenclature of \cite{Cordova:2016emh}
\begin{equation}\label{free3d}
{B_1}[0]_{1/2}^{(1/2,0)}\oplus {B_1}[0]_{1/2}^{(0,1/2)}~.
\end{equation}
The first representation is a free hypermultiplet, and the second representation is a free twisted hypermultiplet. 

\bigskip
\noindent
{\bf Note:} Our conventions differ slightly compared with \cite{Cordova:2016emh}. Here the quantity in square brackets refers to the spin of the primary (not the Dynkin label). The superscripts refer to the spins under the $SO(4)_R\cong SU(2)_R\times SU(2)_C$ $R$-symmetry of the theory (not the Dynkin labels).

\bigskip
\noindent
In the case of \eqref{free3d}, the free hypermultiplet has $R=1/2$ and $C=0$ ($SU(2)_C$ spin zero) while the twisted hypermultiplet has $R=0$ and $C=1/2$. These representations are exchanged by 3d mirror symmetry. Finally, \lq\lq$B_1$" refers to the multiplet shortening condition under the action of supercharges (as in \cite{Cordova:2016emh})
\begin{equation}
Q_{\alpha}^{i,j}\in [1/2]^{(1/2,1/2)}_{1/2}~,
\end{equation}
where $i,j=\pm1$ are spin-half indices of $SU(2)_R$ and $SU(2)_C$ respectively. In particular,
\begin{equation}
\CO^{i_1,0}\in B_1[0]_{1/2}^{1/2,0}\ \Leftrightarrow\ Q^{1,1}_{\alpha}\CO^{1,0}=0~,\ \ \ \tilde\CO^{0,j_1}\in B_1[0]_{1/2}^{0,1/2}\ \Leftrightarrow\ Q^{1,1}_{\alpha}\tilde\CO^{0,1}=0~,
\end{equation}
where the shortening condition is applied to highest-$SO(4)_R$ weight superconformal primaries.

Since mirror symmetry requires us to have free-field defining representations of both $SU(2)_R$ and $SU(2)_C$, and, noting that Euclidean $SU(2)$ spin has less structure than the $R$ symmetry does, it is intuitively clear that we should be able to construct all possible short SCA representations from free fields and their higher-spin currents.

Let us demonstrate this statement explicitly. To that end, note that we can construct any $B_1[0]^{(R,C)}_{{1\over2}(R+C)}$ chiral representation as follows
\begin{eqnarray}
B_1[0]^{(R,C)}_{R+C}&\in& \left(B_1[0]_{1/2}^{(1/2,0)}\right)^{\otimes 2R}\otimes \left(B_1[0]_{1/2}^{(0,1/2)}\right)^{\otimes 2C}~, \cr \CO^{1\cdots1,1\cdots,1}&=& (\CO^{1,0})^{2R}(\tilde\CO^{0,1})^{2C}~,
\end{eqnarray}
where the second line contains highest-weight superconformal primaries of the multiplets on the previous line. In fact, all chiral representations of the 3d $\CN=4$ SCA have $j=0$ \cite{Cordova:2016emh}. Therefore, we have demonstrated that all possible multiplets satisfying $B_1$ shortening conditions can be constructed from free fields. 

Let us consider the other short representations of the 3d $\CN=4$ SCA. In a local theory, we also have a stress-tensor multiplet. For the case of a free hypermultiplet, the primary takes the form
\begin{equation}
J:=\epsilon_{ij}\CO^{i,0}\bar\CO^{j,0}~,\ \ \ (Q^{11})^2J=0\ \Leftrightarrow\ J\in A_2[0]^{(0,0)}_1~,
\end{equation}
where \lq\lq$A_2$" refers to the (quadratic) shortening condition above. In any (local) 3d $\CN=4$ SCFT, the stress tensor multiplet is of type $A_2[0]^{(0,0)}_1$.

More generally, we can use the free stress tensor multiplet to construct any $A_2[0]^{(R,C)}_{1+R+C}$ representation as follows
\begin{eqnarray}
A_2[0]^{(R,C)}_{1+R+C}&\in& A_2[0]^{(0,0)}_{1}\otimes\left(B_1[0]_{1/2}^{(1/2,0)}\right)^{\otimes 2R}\otimes \left(B_1[0]_{1/2}^{(0,1/2)}\right)^{\otimes 2C}~, \cr J^{1\cdots1,1\cdots,1}&=& J(\CO^{1,0})^{2R}(\tilde\CO^{0,1})^{2C}~,
\end{eqnarray}
where we have used the normal-ordered product of the stress tensor multiplet with arbitrary numbers of hypermultiplets and twisted-hypermultiplets. As in the chiral case, all multiplets obeying these shortening conditions have $j=0$ \cite{Cordova:2016emh}. As a result, we have shown that all possible SCA representations satisfying $A_2$ shortening conditions can be built from free fields.

Finally, let us consider the remaining short representations: the $A_1[j]_{1+j+R+C}^{(R,C)}$ multiplets. In a free theory, we have higher-spin currents. These sit in $A_1[j]_{1+j}^{(0,0)}$ multiplets for all $j\ge1/2$ \cite{Cordova:2016emh}. These multiplets allow us to cover the space of possible spins in the SCA, while the free hyper and the free twisted hyper allow us to cover the space of possible $R$ charges. Indeed, we can construct
\begin{eqnarray}
A_1[j]^{(R,C)}_{1+j+R+C}&\in& A_1[j]^{(0,0)}_{1+{j}}\otimes\left(B_1[0]_{1/2}^{(1/2,0)}\right)^{\otimes 2R}\otimes \left(B_1[0]_{1/2}^{(0,1/2)}\right)^{\otimes 2C}~, \cr \mathcal{J}^{1\cdots1,1\cdots,1}_{\alpha_1\cdots\alpha_{2j}}&=& J_{\alpha_1\cdots\alpha_{2j}}(\CO^{1,0})^{2R}(\tilde\CO^{0,1})^{2C}~,
\end{eqnarray}
where $J_{\alpha_1\cdots\alpha_{2j}}$ is the primary of the higher-spin current multiplet.

In summary, we have learned that, in 3d, mirror symmetry, the relative simplicity of Euclidean spin compared to $SO(4)_R$, and the fact that short conformal dimensions are fixed in terms of non-abelian charges conspire to allow us to construct any short multiplet from free fields:

\bigskip
\noindent
{\bf Theorem 2:} Any short representation of the 3d $\CN=4$ SCA can be constructed from free fields.

\bigskip
Let us now consider long multiplets. These transform in representations of the form $L[j]_{\Delta}^{(R,C)}$ with $\Delta>1+j+R+C$. Clearly, we can use free fields to construct a long multiplet with any such $R$, $C$, and $j$ quantum numbers. For example, we can take the above short representations and fuse them with the long multiplet, $J^2:=L[0]^{(0,0)}_2$, obtained from squaring the free stress tensor primary. Of course, we will be unable to reproduce $\Delta\not\in {1\over2}\mathbb{Z}$ in a free theory, but we conjecture that RG flows from free theories lead to anomalous dimensions that allow for all $\Delta$ consistent with unitarity to be constructed. This discussion builds on the intuition we began this section with: Lagrangians and their associated free fields are \lq\lq well-connected" to interacting 3d $\CN=4$ SCFTs. We conjecture that they are connected to any interacting 3d $\CN=4$ SCFT via RG flow.

\newsec{4d and a web of conjectures}
Now we move on to the more interesting case of 4d $\CN=2$ SCFTs. Let us begin by analyzing the half-BPS multiplets of the 4d $\CN=2$ SCA and asking if they admit free field realizations.

First consider $L\bar B_1[j,0]^{(0,r)}_{r}\cong\bar\CE_{r(j,0)}$ multiplets.\footnote{For convenience, in this section we use the notation of both \cite{Cordova:2016emh} and \cite{Dolan:2002zh}. In writing $L\bar B_1[j,0]^{(0,r)}_{r}$, the superscripts indicate $SU(2)_R$ spin (zero in this case) and $U(1)_r$ charge, while the square brackets denote left and right Lorentz spin, and the subscript denotes scaling dimension.} The corresponding primaries satisfy
\begin{equation}\label{barEcond}
\bar Q^i_{\dot\alpha}\CO_{\alpha_1\cdots\alpha_{2j}}=0~.
\end{equation}

\bigskip
\noindent
{\bf Note:} As in 3d $\CN=4$, our $4d$ $\CN=2$ conventions differ from those in \cite{Cordova:2016emh}: $SU(2)_R$ and Lorentz Dynkin labels in \cite{Cordova:2016emh} are replaced by spins here.

\bigskip
\noindent

\begin{table}[h]
    \centering
    \begin{tabular}{c| c c c c c}
        $$ & $R$ & $r$ & $j$ & $E$ & $\delta$ \\ \hline
        $\phi_i$ & 0 & 1 & 0 & 1 & 0 \\ 
        $\lambda^1_{i,\alpha}$ & 1/2 & 1/2 & $\pm$1/2 & 3/2 & 0 \\ 
        $q_a$ & 1/2 & 0 & 0 & 1 & 0 \\ 
        $\tilde{q}_a$ & 1/2 & 0 & 0 & 1 & 0 \\ 
    \end{tabular}
    \caption{Chiral fields in a 4d $\CN=2$ SCFT with $N$ abelian free vector multiplets ($i=1,\cdots,N$) and $M$ free hypermultiplets ($a=1,\cdots,M$). Here $\delta:=E-2R-2\bar j-r$.}
    \label{FreeFields}
\end{table}

In the case of $j=0$, vevs for the primaries in \eqref{barEcond} describe the Coulomb branch. To construct primaries for such representations from free fields, we can only use fields in Table \eqref{FreeFields} that are annihilated by all the $\bar Q^i_{\dot\alpha}$ supercharges. Moreover, we cannot involve derivatives because otherwise we would have $\CO_{\alpha_1\cdots\alpha_{2j}}=\bar Q^i_{\dot\alpha}\CO^{\dot\alpha'}_{i\alpha_1\cdots\alpha_{2j}}$ for some well-defined $\CO^{\dot\alpha'}_{i\alpha_1\cdots\alpha_{2j}}$ and some $i=1,2$. Therefore, any such primary must be built out of the following kinds of products
\begin{equation}
\CO=\prod_i\phi_i^{n_i}~.
\end{equation}
In particular, we can only produce scalar $L\bar B_1[0,0]^{(0,r)}_{r}\cong\bar\CE_{r(0,0)}:=\bar\CE_{r}$ representations with $r\in\mathbb{Z}$. Since higher-spin $\bar\CE_{r(j,0)}$ with   $j\neq 0$ multiplets are not related to the ones realized in free theories by a continuous deformation (spin is quantized), the intuition in the introduction suggests that such multiplets are absent in general local unitary interacting 4d $\CN=2$ SCFTs.

This claim turns out to be correct. For example, in \cite{Buican:2014qla} it was shown that various classes of theories cannot have $j>0$  $L\bar B_1[j,0]^{(0,r)}_{r}$ multiplets. Finally, in \cite{Manenti:2019kbl}, it was shown that only scalar $L\bar B_1[0,0]^{(0,r)}_{r}$ multiplets can exist in local 4d $\CN=2$ SCFTs. The reason is that the three-point functions of these multiplets, their conjugates, and the supercurrent multiplet only satisfy the superconformal Ward identities when $j=0$.

Therefore, the $\bar\CE_{r}$ representations that exist in local unitary interacting SCFTs are precisely those that are connected to the $\bar\CE_r$ representations that exist in free theories by a continuous deformation (of $U(1)_r$). This result is compatible with the intuition that theories with the most general $\bar\CE_r$ representations are connected to free theories via a continuous deformation in the space of theories. Indeed, all known local unitary interacting 4d $\CN=2$ SCFTs have a Coulomb branch. Therefore, in such theories, we expect that turning on generic vevs for $\bar\CE_r$ primaries and/or turning on generic relevant prepotential deformations takes us to a free theory. We then have the following map of chiral rings
\begin{equation}\label{Ering}
\mathcal{R}_{UV}:=\langle\bar\CE_{r_1}~,\ \bar\CE_{r_2}~,\ \cdots~,\ \bar\CE_{r_N} \rangle\xrightarrow{\rm RG}\mathcal{R}_{IR}:=\langle\bar\CD_{0(0,0)}^{(1)}~,\ \bar\CD_{0(0,0)}^{(2)}~,\ \cdots~,\ \bar\CD_{0(0,0)}^{(N)} \rangle~,
\end{equation}
where the brackets indicate that the ring is generated by the primaries of the enclosed multiplets, $N$ is the rank of the theory (i.e., the dimension of the Coulomb branch or the number of independent $\bar\CE_r$ generators), and $\bar\CD_{0(0,0)}^{(i)}$ is the $i$-th free vector multiplet (for simplicity, we can assume the Coulomb branch chiral ring is freely generated).\footnote{Moreover, under dimensional reduction, these multiplets typically flow to monopole operators corresponding to highest-$SU(2)_C$ weight elements of $B_1[0]^{(0,C)}_{C}$ multiplets \cite{Buican:2015hsa}. As discussed in the previous subsection, these degrees of freedom admit a free field representation in 3d and are often related to UV Lagrangians.}

Next we consider the so-called \lq\lq Schur sector" of the SCA. These states are related to 2d VOAs via the map in \cite{Beem:2013sza}. Let us begin by considering the $B_1\bar B_1[0,0]_{2R}^{(R,0)}\cong\hat\CB_R$ multiplets whose vevs describe Higgs branches (these are elements of the so-called \lq\lq Hall-Littlewood" subring of the Schur sector). The primaries satisfy
\begin{equation}
\bar Q^1_{\dot\alpha}\CO^{1\cdots1}=Q^1_{\alpha}\CO^{1\cdots1}=0~.
\end{equation}
These representations are related to the 3d $\CN=4$ $B_1[0]_{R/2}^{(R)}$ multiplets described in the previous subsection via dimensional reduction. As in that case, the free hypermultiplet is in the defining representation of $SU(2)_R$ and therefore generates all multiplets in the family
\begin{equation}
B_1\bar B_1[0,0]_{2R}^{(R,0)}\in \left(B_1\bar B_1[0,0]_1^{(1/2,0)}\right)^{\otimes 2R}~,\ \ \ \CO^{1\cdots1}=(\CO^1)^{2R}=q^{2R}~.
\end{equation}
Therefore, any $\hat\CB_R$ representation admits a free field representation in terms of free hypers (which sit in the $\hat \CB_{\frac12}$ multiplet). This fact is compatible with the known non-renormalization theorems for such moduli spaces (the structure of the corresponding ring does not change under RG flow to generic free points on the Higgs branch) \cite{Argyres:1996eh}.

\begin{table}[]
    \centering
    \begin{tabular}{c| c c c c c c}
        $$ & $R$ & $r$ & $j$ & $\bar j$ & $E$ & $\delta$ \\ \hline
        $\lambda^1_{i,+}$ & 1/2 & 1/2 & 1/2 &0 & 3/2 & 0 \\
        $\bar \lambda^1_{i,\dot+}$ & 1/2 & -1/2 & 0 & 1/2 & 3/2 & 0 \\ 
        $q_a$ & 1/2 & 0 & 0& 0 & 1 & 0 \\ 
        $\tilde{q}_a$ & 1/2 & 0 & 0 &0 & 1 & 0 \\
        $\partial_{+\dot+}$ &0 & 0 &1/2 & 1/2 & 1 & 0
    \end{tabular}
    \caption{Schur fields (and derivative) in a 4d $\CN=2$ SCFT with $N$ abelian free vector multiplets ($i=1,\cdots,N$) and $M$ free hypermultiplets ($a=1,\cdots,M$). Here $\delta:=E-2R-2\bar j-r$.}
    \label{FreeFieldsSchur}
\end{table}

Next let us consider $\bar\CD_{R(j,0)}$ Schur multiplets. A simple result in \cite{Banerjee:2023ddh} shows that:

\bigskip
\noindent
{\bf Theorem 3 \cite{Banerjee:2023ddh}:} In free 4d $\CN=2$ SCFTs, $A_a\bar{B}_1[j,0]_{1+j}^{(R,1+j)}\cong\bar\CD_{R(j,0)}$ multiplets have $j\le R$.

\bigskip
\noindent
In more general local unitary theories the same condition holds as a consequence of the ANEC \cite{Manenti:2019kbl}. This statement is again compatible with our discussion, as we expect these multiplets to be related via RG flow in a similar manner to those in \eqref{Ering}. 

What about the remaining Schur operators? They live in $A_a\bar A_b[j,\bar j]^{(R,(j-\bar j)/2)}\cong\hat\CC_{R(j,\bar j)}$ multiplets and satisfy the following shortening conditions for their primaries
\begin{eqnarray}
Q^{1\alpha_1}\CO^{1\cdots1}_{\alpha_1\cdots\alpha_j;\dot\alpha_1\cdots\dot\alpha_{\bar j}}&=&\bar Q^{1\dot\alpha}\CO^{1\cdots1}_{\alpha_1\cdots\alpha_j;\dot\alpha_1\cdots\dot\alpha_{\bar j}}=0~, \ \ \ j,\bar j\ne0~,\cr Q^{1\alpha_1}\CO^{1\cdots1}_{\alpha_1\cdots\alpha_j}&=&(\bar Q^{1})^2\CO^{1\cdots1}_{\alpha_1\cdots\alpha_j}=0~, \ \ \ j\ne0~,\ \bar j=0~,\cr (Q^{1})^2\CO^{1\cdots1}_{\dot\alpha_1\cdots\dot\alpha_{\bar j}}&=&\bar Q^{1\dot\alpha}\CO^{1\cdots1}_{\dot\alpha_1\cdots\dot\alpha_{\bar j}}=0~, \ \ \ \bar j\ne0~,\ j=0~,\cr (Q^{1})^2\CO^{1\cdots1}&=&(\bar Q^{1})^2\CO^{1\cdots1}=0~, \ \ \  j=\bar j=0.
\end{eqnarray}

Given the bounds on the other Schur multiplets arising from locality, it is reasonable there should be a corresponding bound for $\hat\CC$ representations. Indeed, the Schur ring is a closed subsector of any 4d $\CN=2$ SCFT and all other Schur representations are related to $\hat\CC$ multiplets via continuation to negative spin: 
\be\label{CDrelation}
{\hat \cC_{R (-\frac12,\bar j)}}\cong \CD_{R+\frac12(0,\bar j)}~, \qquad
{\hat \cC_{R (j,-\frac12 )}}\cong \bar\CD_{R+\frac12(  j,0)}~, \qquad
{\hat \cC_{R (-\frac12,-\frac12)}}\cong  \hat \CB_{R+1}~. \qquad
\ee
Moreover, it is natural to assume that such a bound including all Schur operators is linear in the superconformal quantum numbers. 

Given these assumptions, we can uniquely fix the form of a general Schur locality bound. To understand this statement, first note that, by CPT invariance of the $\hat\CC_{R(j,\bar j)}$ spectrum, the most general inequality possible takes the form
\begin{equation}\label{genineq}
a_r|r|+a_+(j+\bar j)+a_RR+a\le0~,
\end{equation}
where $r=j-\bar j$. It is easy to argue that $a_+, a_R, a\le0$. Indeed, first consider the implications of the existence of higher-spin currents. On general grounds, these multiplets are of type $\hat\CC_{0(j,\bar j)}$. In a free theory (or a discrete gauging thereof), the corresponding Schur operators are built from the fields and derivative in Table \ref{FreeFieldsSchur}. Since they are currents, these operators are quadratic in fields (but arbitrary order in derivatives). Therefore, we see that the Schur multiplets have spin $(j,j)$, $(j,j-1)$, $(j-1,j)$, $(j-1/2,j)$, and $(j,j-1/2)$. The latter two sets of multiplets correspond to mixed currents involving vector and hypermultiplet degrees of freedom. Since the Schur operators are the $Q^1_+\bar Q^1_{\dot+}$-descendant of the primaries, we see these multiplets transform in the following representations of the SCA
\begin{eqnarray}\label{hs}
A_1\bar A_1[j,j]^{(0,0)}_{2+2j}\cong\hat\CC_{0(j,j)}&\oplus& A_1\bar A_b[k,k-1]^{(0,1)}_{1+2k}\cong\hat\CC_{0(k,k-1)}\cr&\oplus& A_a\bar A_1[k-1,k]^{(0,-1)}_{1+2k}\cong\hat\CC_{0(k-1,k)} \cr&\oplus& A_1\bar A_b[j,j-1/2]_{3/2+2j}^{(0,1/2)}\cong\hat\CC_{0(j,j-1/2)}\cr&\oplus& A_a\bar A_1[j-1/2,j]_{3/2+2j}^{(0,-1/2)}\cong\hat\CC_{0(j-1/2,j)}~,\ j\in \mathbb{N}/2~,\ k\in \mathbb{N}~.
\end{eqnarray}
As a result, we have $a_+\le0$ since these currents can have arbitrarily large $j+\bar j$. Next, note that any local 4d $\CN=2$ SCFT has an energy-momentum tensor sitting in a $\hat\CC_{0(0,0)}$ multiplet. As a result, locality implies $a\le0$. Finally, note that in the case of a theory of a free hypermultiplet, we have $\hat\CC_{R(0,0)}$ representations arising from the normal-ordered products of primaries in $\hat\CB_R\times \hat\CC_{0(0,0)}$. Therefore, since $R$ can be arbitrarily large, we must have $a_R\le0$.

We should also demand consistency of our bound with the known ANEC bound involving $\bar\CD_{R(j,0)}$ multiplets. In this case, we have the map $\bar\CD_{R+1/2(j,0)}\cong\hat\CC_{R(j,-1/2)}$, and $j\le R+1/2$. Clearly, since $a_+, a_R, a\le0$, the only way to obtain a non-trivial inequality from \eqref{genineq} is for $a_r>0$. Without loss of generality, we set $a_r=1$ and obtain
\begin{equation}\label{genineq2}
|r|+a_+(j+\bar j)+a_RR+a\le0~.
\end{equation}
Taking $\bar j=-1/2$, setting $r=1+j$, and substituting into \eqref{genineq2}, we obtain
\begin{equation}
1+j+a_+(j-1/2)+a_R(R+1/2)+a\le0~.
\end{equation}
This inequality must reduce to $j\le R+1/2$ and therefore requires that
\begin{equation}\label{ass}
a_++1=-a_R~,\ \ \ a_++1=2(a_+/2-a-a_R/2-1)\ \Rightarrow\ a=-a_R/2-3/2=a_+/2-1~.
\end{equation}
As a consequence of the first equality, we have that $a_+>-1$.

To get a stronger bound, note that $\hat\CC_{0(j,j+k)}$ cannot exist for $k>1$. This follows from the discussion around \eqref{hs}.\footnote{Another way to rule out such multiplets (at least for $j=0$) is as follows. We note that the ANEC analysis in \cite{Manenti:2019kbl} rules out $A_1\bar L$ $\CN=1$ shortening conditions for
\begin{equation}
\bar q:={1\over2}\left(\Delta-{3\over2}r_{\CN=1}\right)={1\over2}\left(\Delta-r-2R\right)\ge j-{1\over2}~,
\end{equation} 
where the shortening conditions correspond to
\begin{equation}\label{shortA1bA2N1}
Q_1^{\alpha_1}\CO_{\alpha_1\alpha_2\cdots\alpha_{2j}}=(\bar Q^1)^2\CO_{\alpha_1\cdots\alpha_{2j}}=0~.
\end{equation}
These primaries have $r_{\CN=1}=2j/3$, where $r_{\CN=1}={2\over3}(r+2R)$. Let us suppose this is for a $\CC_{R,r(j,0)}$ primary. We have $\Delta=2+2j+2R-r$ and
\begin{equation}
\bar q=1+j-r\ge j-1/2\ \Rightarrow\ r\le 3/2~.
\end{equation}
We also have that $r<j$ from unitarity (when $r=j$, we have a $\hat\CC_{R(j,0)}$ multiplet). Since we satisfy the first bound in \eqref{shortA1bA2N1}, we must have $R=0$, at least if this is a constraint on the highest-weight primary (the $\CN=2$ shortening condition involves $Q^{1\alpha_1}$ contracted with the highest-$SU(2)_R$-weight primary; therefore, imposing the first condition in \eqref{shortA1bA2N1} is achieved through $SU(2)_R$ lowering). Next, consider the level-one descendant, $\CO'_{\alpha\alpha_1\cdots\alpha_{2j}}:=Q^1_{\alpha}\CO_{\alpha_1\alpha_2\cdots\alpha_{2j}}$. It has $j_1=j+1/2$, $r_1=r-1/2$, $R_1=R+1/2=1/2$, and $\Delta_1=\Delta+1/2$. Therefore
\begin{equation}
{1\over2}\left(\Delta_1-{3\over2}r_{1,\CN=1}\right)={1\over2}\left(\Delta_1-r_1-2R_1\right)\ge j_1+{1\over2}\ \Rightarrow\ r_1\le1/2~.
\end{equation}
As a result, we see that any $\CC_{0,r(j,0)}$ multiplet has $r\le0$. This statement is consistent with $\CC_{R,r(j,\bar j)}$ satisfying $\bar j\le j+R+1$ (in the present case, this latter statement leads to $r<j-\bar j\ge-1$). By complex conjugating, we obtain
\begin{equation}
\bar\CC_{0,r(0,\bar j)} \ \Rightarrow\ r\ge0~.
\end{equation}
Now, suppose that $\hat\CC_{0(0,k)}$ exists for $k>1$. Then, fusing the primary with $\phi$ we obtain a $\bar\CC_{0,1-k(0,k)}$ multiplet. This is a contradiction.} In this case, \eqref{genineq} and \eqref{ass} gives
\begin{equation}\label{highk}
k+a_+(2j+k)+a_+/2-1\le0~.
\end{equation}
Now, fix some finite $k>1$ and consider all possible $j$. Then \eqref{highk} should be violated as the corresponding multiplets don't exist. Since $j$ can be arbitrarily large, we see that $a_+=0$ in order for our inequality to rule out the $\hat\CC_{0(j,j+k)}$ multiplets (a negative $a_+$ means \eqref{highk} is always valid for large enough $j$, and thus cannot impose any constraints on $\hat\CC_{0(j,j+k)}$ for large enough spin).

As a result, we arrive at

\bigskip
\noindent
{\bf Conjecture 1:} The spectrum of $\hat\CC_{R(j,\bar j)}$ multiplets in a local unitary 4d $\CN=2$ SCFT satisfies the constraint
\begin{equation}\label{Chatbound}
|r|=|j-\bar j|\le R+1~.
\end{equation}
In fact, this bound applies uniformly to all Schur operators after making the identification in \eqref{CDrelation} (although it is weaker than the unitarity bound for $\hat\CB_R$ representations). In particular, we recover the ANEC bound discussed below theorem 3, namely  $j\le R$ for the $\bar\CD_{R(j,0)} $ multiplets.

\bigskip
\noindent
As in the case of the $\bar\CD_{R(j,0)}$ multiplets, $\hat\CC_{R(j,\bar j)}$ representations built from free fields obey these bounds:

\bigskip
\noindent
{\bf Theorem 4:} In free theories, $\hat\CC_{R(j,\bar j)}$ multiplets satisfy $|r|=|j-\bar j|\le {R}+1$.

\bigskip
\noindent
{\bf Proof:} In a free theory, the Schur operators are generated by $\partial_{+\dot+}$, the $N$ gaugino pairs, $\lambda^1_{i,+}$ and $\bar\lambda^1_{i,\dot+}$, and the $M$ hypermultiplet scalars, $q_a$ and $\tilde q_a$ (see Table \ref{FreeFieldsSchur}). Clearly, such operators are highest weight with respect to spin and $SU(2)_R$. Since each of these letters has $|j-\bar j|\le R$, we see that the $\hat\CC$ primary has $|j-\bar j|\le R+1$. $\square$

\bigskip
Schur operators are also known to correspond to states in a 2d VOA \cite{Beem:2013sza}. It is therefore interesting to ask if this correspondence can shed light on our conjecture.

To that end, let us begin with another reasonable conjecture on the spectrum of such VOAs. Namely, when there is a 4d Higgs branch, the 2d images of the 4d Hall-Littlewood (HL) operators form a set of generators of the chiral algebra \cite{Beem:2019tfp}. Note that these are not strong generators in general, so we should also include singular terms in the corresponding OPEs in order to obtain the full set of VOA states.

Let us study theories for which the images of HL operators generate the VOA and ask if \eqref{Chatbound} is satisfied in that case. For simplicity, we will stick to the case where the HL multiplets have $r=0$, so that the generators sit in $\hat\CB_R$ multiplets.\footnote{Examples of such VOAs include those in \cite{Buican:2015ina,Buican:2017fiq}.}

To answer this question, we should find all possible Schur operators that appear in the OPEs of the Schur operators (and their $SU(2)_R$ relatives) in $\hat\CB_{R_1}\times\hat\CB_{R_2}$. In any of these OPEs, $U(1)_r$ covariance dictates that the righthand sides of such OPEs have the same $U(1)_r$ charge and must therefore have $r=0$. In the case of $\hat\CC_{R(j,\bar j)}$ multiplets, both the primary and the Schur operators have the same $r=j-\bar j$ (this is because the Schur operator is the $Q^1_+\bar Q^1_{\dot+}$ descendant). Therefore, any $\hat\CC_{R(j,\bar j)}$ multiplets appearing in the OPEs have $|r|=|j-\bar j|=0$ and so \eqref{Chatbound} is satisfied. Similar logic shows that for any further OPEs in the theory, the corresponding Schur operators will all have $r=0$ and our conjecture is satisfied.\footnote{This discussion has immediate implications for class $\CS$ theories with regular punctures \cite{Gaiotto:2009we}. Many such theories can be constructed by starting from copies of the 6d $(2,0)$ theory compactified on spheres with three punctures. Associated with each puncture is a flavor symmetry in the resulting 4d $\CN=2$ SCFT. Such \lq\lq trinion" theories can then be glued together via gauging the symmetries associated with these punctures. Through this procedure, we can obtain class $\CS$ theories corresponding to compactifications of the $(2,0)$ theory on surfaces of any genus.

How can the above discussion be translated into this context? First, it is widely believed that the HL sectors of the trinion theories are built purely from $\hat\CB_R$ multiplets. Since these multiplets have $r=0$, the discussion below Theorem 4 implies that the corresponding Schur sector satisfies our bounds (if the conjecture in \cite{Beem:2019tfp} holds). However, we can say more. Indeed, gluing in the class $\CS$ context corresponds to gauging via the addition of vector multiplets and the construction of gauge-invariant operators. The corresponding vector multiplet Schur operators are gauginos satisfying $j(\lambda)=R(\lambda)$ (or $\bar j(\bar\lambda)=R(\bar\lambda)$). As a result, composite Schur operators built from gluing together trinions obey our bounds as well.} More generally, we believe any Schur sector satisfies our conjecture, but it is not immediately clear how to use the VOA map itself to prove it.

\bigskip
Next let us consider the $\bar\CB_{R,r(j,0)}$ representations. First, we have the result from \cite{Banerjee:2023ddh}

\bigskip
\noindent
{\bf Theorem 5 \cite{Banerjee:2023ddh}:} In free 4d $\CN=2$ SCFTs, $\bar\CB_{R,r(j,0)}$ multiplets have $j\le R$.

\bigskip
\noindent
This prompted the following conjecture:

\bigskip
\noindent
{\bf Conjecture 2 \cite{Banerjee:2023ddh}:} In any local unitary 4d $\CN=2$ SCFT, $\bar\CB_{R,r(j,0)}$ multiplets have $j\le R$.

\bigskip
\noindent
One additional piece of evidence given in \cite{Banerjee:2023ddh} was that, for $r<j+2$, the ANEC results of \cite{Manenti:2019kbl} imply the above conjecture is true.

Let us give some additional evidence in favor of the above conjecture. To that end, in all theories we are aware of: either {\bf(1)} $\bar\CB_{R,r(j,0)}$ multiplets are composites of other multiplets (in particular, the chiral operators in $\bar\CB_{R,r(j,0)}$ multiplets are composites of chiral operators in other types of multiplets; see \cite{Banerjee:2023ddh} for some explicit examples); or {\bf(2)} $\bar\CB_{R,r(j,0)}$ chiral generators arise in the following way when we gauge some continuous $G$ symmetry (exactly marginally or not) of an $\CN=2$ theory
\begin{equation}\label{TrGenN2}
\tilde\CO:={\rm Tr}\left(\phi^n\prod_i\CO_i\right)\in\bar\CB_{R,r(j,0)}~,
\end{equation}
where the trace renders $\tilde\CO$ invariant under $G$, $\phi$ is a $G$ vector multiplet (we can include $\delta=0$ holomorphic gauginos here as well), and $\CO_i\in\hat\CB_R$, $\CO_i\in\bar\CB_{R,r(j,0)}$ (these latter operators are generated as in {\bf(1)}), and $\CO_i\in\bar\CD_{R(j,0)}$ are chiral operators transforming in some representation of $G$.\footnote{Recall that $\bar\CE_r$ multiplets are invariant under $G$ \cite{Buican:2013ica,Buican:2014qla}.} Relatedly, through discrete gauging, we could consider $\bar\CB_{R,r(j,0)}$ generators arising from invariant products of operators in $\hat\CB_R$, $\bar\CB_{R,r(j,0)}$, and $\bar\CD_{R(j,0)}$ transforming under a discrete gauge group.\footnote{However, we have not systematically checked that such putative examples are free of obstructions to discrete gauging from 't Hooft anomalies.}

In the case of scenario {\bf(1)}, we can prove the following theorem:

\bigskip
\noindent
{\bf Theorem 6:} If the chiral ring of a 4d $\CN=2$ SCFT, $\CT$, is generated by operators in $\bar\CE$, $\hat\CB$, and $\bar\CD$ multiplets (or a subset thereof), then all $\bar\CB$ multiplets in $\CT$ have $j\le R$.

\bigskip
\noindent
{\bf Proof:} As shown in \cite{Bhargava:2022cuf}, any chiral operator in an $\CN=2$ SCFT must sit as either a superconformal primary, a $Q^1_{\alpha}$ descendant, or a $(Q^1)^2$ descendant of an $\bar\CE_r$, $\hat\CB_R$, $\bar\CD_{R(j,0)}$, or $\bar\CB_{R,r(j,0)}$ multiplet. These chiral operators are always $SU(2)_R$ highest weight within their $SU(2)_R$ representations. We know that the first three types of multiplets have primaries satisfying $j\le R$ \cite{Manenti:2019kbl,Manenti:2019jds}. Since $j(Q^1_{\alpha})\le R(Q^1_{\alpha})$, any chiral operator in a $\bar\CE_r$, $\hat\CB_R$, and $\bar\CD_{R(j,0)}$ multiplet has $j\le R$. By assumption, products of such operators give the chiral operators in any $\bar\CB_{R,r(j,0)}$ multiplet. As a result, these latter chiral operators have $j\le R$, and all $\bar\CB_{R,r(j,0)}$ multiplets satisfy $j\le R$ as claimed. $\square$

Using a similar argument in the case of {\bf(2)}, we arrive at the following statement:

\bigskip\bigskip
\noindent
{\bf Theorem 7:} $\bar\CB_{R,r(j,0)}$ multiplets generated via the process in \eqref{TrGenN2} have $j\le R$.\footnote{We could consider similar operators arising from products of operators in $\hat\CB_R$, $\bar\CB_{R,r(j,0)}$, and $\bar\CD_{R(j,0)}$ via discrete gauging. Our conclusion is unchanged in this case.}

\bigskip
\noindent
More generally, we could imagine constructing operators as in \eqref{TrGenN2} but starting with some $\CN=1$ theory that then enhances via RG flow to an $\CN=2$ theory in the IR. If the UV theory is an asymptotically free $\CN=1$ gauge theory, then the only $\CN=1$ chiral fields with spins are gauginos, $\lambda^{A_i}_{i,\alpha}$ (where $i$ denotes the $i$th gauge group, and $A_i$ is a corresponding adjoint index). In order to be a useful Lagrangian in the spirit of \cite{Maruyoshi:2016tqk}, both $r$ and $R$ should be visible in the RG flow to the IR. Therefore, we have $j(\lambda^{A_i}_{i,\alpha})=r(\lambda^{A_i}_{i,\alpha})=R(\lambda^{A_i}_{i,\alpha})=1/2$. As long as all the $\CN=1$ chiral primaries have $R\ge0$, these theories clearly only have $\bar\CB_{R,r(j,0)}$ multiplets satisfying $j\le R$.\footnote{We thank A.~Banerjee for discussions on this point.} Examples of such theories were analysed in \cite{Banerjee:2023ddh}. More general $\CN=1$ Lagrangians may violate $R\ge0$ for all the chiral primaries (e.g., see \cite{Gadde:2015xta}), but, even in these cases, we are not aware of the existence of $\bar\CB_{R(j,0)}$ multiplets with $j>R$.

Next, let us consider the $\bar\CC_{R,r(j,\bar j)}$ multiplets (our results below imply the corresponding $\bar\CB$ results above after identifying $\bar \cC_{R (j,-\frac12 )}\cong \bar\CB_{R+\frac12,r-\frac12(  j,0)}$). In free theories, we have the following result:

\bigskip
\noindent
{\bf Theorem 8:} In free 4d $\CN=2$ SCFTs, $\bar\CC_{R,r(j,\bar j)}$ multiplets satisfy\footnote{As a consequence, $\CC_{R,r(j,\bar j)}$ multiplets satisfy
\begin{equation}
\bar j\le j+R+1~.
\end{equation}
}
\begin{equation}
j\le \bar j+R+1~.
\end{equation}

\bigskip
\noindent
{\bf Proof:} The $\delta=0$ states in the $\bar\CC$ multiplet are generated by $\phi_i$, $\lambda^1_{i,\alpha}$, $\bar\lambda^1_{i,\dot+}$, $\bar F_{i,\dot+\dot+}$, $\partial_{\alpha\dot+}$, $q_a$, $\tilde q_a$, and $\bar\psi_{a,\dot+}$, and $\bar{\tilde{\psi}}_{a,\dot+}$. Here $\delta:=E-2R-2\bar j-r$. Each of these operators satisfies $j\le \bar j+R$. The $\delta=0$ states in a $\bar\CC$ multiplet come from acting on the $SU(2)_R$ and $SU(2)_{\bar j}$ highest-weight $\CN=2$ superconformal primary with $\bar Q^1_{\dot+}$ and at most one of each of $Q^1_{\alpha}$, $\bar Q^1_{\dot-}$, and $\bar Q^2_{\dot+}$. Acting with $\bar Q^1_{\dot-}$ and $\bar Q^2_{\dot+}$ does not change $\bar j+R$ or $j$. On the other hand, acting with $\bar Q^1_{\dot+}$ raises $\bar j+R$ by one unit and forces the primary to satisfy $j\le\bar j+R+1$. Acting with $Q^1_{\alpha}$ gives $j\to j\pm1/2$ in addition to raising $R\to R+1/2$. Acting with $(Q^1)^2$ takes $R\to R+1$ but leaves the other quantum numbers unchanged. As a result, we conclude that $j\le \bar j+R+1$ is the strongest constraint on the primary. $\square$

\bigskip
\noindent
As in the cases of other multiplets, the above discussion leads us to the following conjecture

\bigskip
\noindent
{\bf Conjecture 3}:  In any local unitary 4d $\CN=2$ SCFT, $\bar\CC_{R,r(j,\bar j)}$ multiplets satisfy:
\begin{equation}\label{Cconj}
j\le \bar j+R+1~.
\end{equation}

\bigskip
\noindent
The additional evidence in favor of the above conjecture is similar in spirit to that for the $\bar\CB_{R,r(j,0)}$ conjecture. Note that, in what follows, we will assume the $\bar\CB_{R,r(j,0)}$ and $\hat\CC_{R(j,\bar j)}$ conjectures hold. Indeed, in all theories we are aware of these conjectures hold, and: either {\bf(1)} $\bar\CC_{R,r(j,\bar j)}$ are composites of other multiplets (in particular, this holds for the $\delta=0$ states); {\bf(2)} $\bar\CC_{R,r(j,\bar j)}$ $\delta=0$ generators arise from gauging a continuous $G$ symmetry (exactly marginal or not) of an $\CN=2$ theory; or {\bf(3)} $\bar\CC_{R,r(j,\bar j)}$ generators arise from an $\CN=1\to\CN=2$ enhancing RG flow.

Let us consider possibility {\bf(1)} first. We have

\bigskip
\noindent
{\bf Theorem 9:} If the $\bar\CB_{R,r(j,0)}$ representations satisfy $j\le R$, and the $\delta=0$ states of a $\bar\CC_{R,r(j,\bar j)}$ multiplet involve normal-ordered products of operators in other types of multiplets, then we have $j\le \bar j+R+1$.

\bigskip
\noindent
{\bf Proof:} Such a $\delta=0$ state must involve $\delta=0$ states from a $\hat\CC$ multiplet and arbitrarily many states from the chiral sector (i.e., $\bar\CE_r$, $\hat\CB_R$, $\bar\CD_{R(j,0)}$, and $\bar\CB_{R,r(j,0)}$). The $\delta=0$ states of $\hat\CC$ can be reached from the highest $SU(2)_R$ and Lorentz-weight superconformal primary of $\hat\CC$ and acting with $\bar Q^1_{\dot+}$ and at most one of $\bar Q^2_{\dot+}$, $Q^1_+$, and $Q^1_-$. It is easy to check that each of these $\delta=0$ states satisfies $j'\le \bar j'+R'$ (where these are the $j$, $\bar j$, and $R$ quantum numbers of these states). As we saw in the proof of the theorem on composite $\bar\CB$ multiplets, all chiral states in chiral multiplets satisfy $j'\le R'$ (and $\bar j'=0$). As a result, we have that $\bar\CC_{R,r(j,\bar j)}$ satisfies $j\le\bar j+R+1$. $\square$

\bigskip
\noindent
In the case of {\bf(2)}, we construct $\delta=0$ states via
\begin{equation}\label{CgenN2}
\tilde\CO:={\rm Tr}\left(\phi^n\prod_i\CO_i\right)\in\bar\CC_{R,r(j,\bar j)}~,
\end{equation}
where $\phi$ is a $G$ vector multiplet (we can include $\delta=0$ gauginos here as well), $\CO_i\in\hat\CC_{R'(j',\bar j')}$, or $\CO_i$ is a chiral operator in a chiral multiplet (we can also allow for $\delta=0$ states in $\bar\CC$ multiplets satisfying {\bf(1)}). Here the trace renders $\tilde\CO$ invariant under $G$. As in the proof of the previous theorem, we have

\bigskip
\noindent
{\bf Theorem 10:} $\bar\CC_{R,r(j,\bar j)}$ multiplets generated via the process in \eqref{CgenN2} have $j\le \bar j+R+1$.

\bigskip
\noindent
Finally, we can consider case {\bf(3)}. Here we simply note that all examples we have checked are consistent with our conjecture (e.g., this holds for all examples considered in \cite{Banerjee:2023ddh}).

Heuristically, we can give the following additional arguments for the above bounds. Up to the existence of null states that do not follow from statistics, chiral operators behave very similarly to free fields. Indeed, consider chiral operators $\CO_{1,2}$ with quantum numbers $(E_i, R_i, r_i, j_i)$, where $i=1,2$. In the absence of null states, we have an operator $\CO_{12}:=\CO_1\CO_2$ with quantum numbers $(E_1+E_2, R_1+R_2, r_1+r_2, j_1+j_2)$. Therefore, it is natural that such operators (in representations $\bar\CE_r$, $\hat\CB_R$, $\bar\CD_{R(j,0)}$, and $\bar\CB_{R,r(j,0)}$) should satisfy bounds given by free fields (null relations should not change the picture since these simply remove operators from the spectrum).

For the case of the $\hat\CC_{R(j, \bar j)}$ and $\bar\CC_{R,r(j,\bar j)}$ multiplets, a heuristic argument in favor of the bounds is as follows. For sufficiently large total spin (i.e., $j+\bar j$), aspects of the spectrum of any $d>2$ CFT resemble aspects of the spectrum of (generalized) free fields \cite{Komargodski:2012ek,Fitzpatrick:2012yx} (in particular the additivity of twists, much like the additivity of conformal dimensions in the chiral case). Moreover, for $j+\bar j\gg R$, we can naively \lq\lq forget" about $R$. In this way, we see the heuristic emergence of \lq\lq effective" $\hat\CC_{0(j,\bar j)}$ multiplets, and the constraints that follow from free fields. On the other hand, for $j+\bar j\lesssim R$, we expect $|j-\bar j|\lesssim R$ and for our bounds to hold.

\subsec{Implications for the index, flavor symmetries, and VOAs}
In this section, we begin by looking at the implications for the superconformal index assuming the above conjectures hold. To that end, recall that the superconformal index is defined as follows
\begin{equation}
\CI(p,q,t):={\rm Tr}|_{\CH}(-1)^Fq^{j+\bar j+r}p^{\bar j-j+r}t^{R-r}~,
\end{equation}
where the trace is over the space of local operators, $\CH$, $(-1)^F$ is fermion number, and $p$, $q$, $t$ are fugacities for the corresponding superconformal charges. Unitarity implies that $\bar j\pm j+r\ge0$, but $R-r$ does not have a definite sign. The index can also be additionally refined by flavor fugacities, but we leave such quantities implicit in our discussion. It turns out that the only operators contributing to the index satisfy
\begin{equation}
\delta:=E-2R-2\bar j-r=0~.
\end{equation}

To make contact with our conjectures it is useful to introduce $u:=pq/t$ and rewrite the index in terms of $q$, $u$, and $t$
\begin{equation}
\CI(q,u,t)={\rm Tr}|_{\CH}(-1)^Fq^{2j}u^{\bar j-j+r}t^{R+\bar j-j}~.
\end{equation}
From our discussion in the previous section, we see that, assuming our conjectures hold,  $R+\bar j-j\ge0$ for all states with $\delta=0$.\footnote{For example, in the $\bar \cC_{R,r(j,\bar j)}$ multiplet, the conjecture \eqref{Cconj} gives the constraint  $R+\bar j-j+1\ge 0$ on the superconformal primary operator. Acting with $\bar Q^1_{\dot +}$ gives the level-1 operator  $X$ with $\delta_X=0$ and quantum numbers $[R_X,r_X,j_X,\bar j_X]=[R+\frac12, r-\frac12, j,\bar j+\frac12]$ which is subject to the condition $R_X+\bar j_X-j_X\ge0$. Similar conclusions can be reached for other $\delta=0$ operators in $\bar \cC$ multiplets as well as  $\hat \cC$ multiplets.}
Therefore, we can in principle take the following limit 
\be\label{GCBlimit}
t,p \to 0~,\qquad \text{with } q,u  \text{ fixed}~,
\ee
in the index without worrying about divergences (this limit has been defined and discussed in \cite{Gadde:2011uv}). Note that this statement is rigorously true for $\bar\CD$ multiplets by the ANEC results described above.

If we take the above limit in a theory of $N$ free abelian vector multiplets and $M$ free hypermultiplets, the only contributing states are
\begin{equation}\label{GCff}
\phi_i~,\ \lambda^1_{i,+}~,\ \partial_{+\dot+}~,
\end{equation}
and so the index reduces as follows
\begin{equation}
\CI_{GC}(q,u)=\prod_i{\rm P.E.}\left({u-q\over1-q}\right)~.
\end{equation}
Since the contributions only come from elements of the $\bar\CD_{0(0,0)}$ gauge multiplet (note that the hypermultiplets do not contribute), we refer to this limit as the \lq\lq generalized Coulomb" limit.

The above discussion has interesting connections with other limits of the index. For example, taking $q\to0$ yields the Coulomb limit (only the $\phi_i$ contribute). On the other hand, taking $u\to0$ yields a $t\to0$ limit of the Macdonald index which we will call the \lq\lq reduced" Macdonald limit.\footnote{We have checked that, in the absence of additional assumptions, there are no other special limits of the index arising from taking subsets of the fugacities to vanish.} In a free theory, we only get contributions from $\lambda^1_{i,+}$ and $\partial_{+\dot+}$. Since this is not enough to give us an $SU(2)_R$ current (for that, we need $\bar\lambda_{\dot+}^1$ as well), in the 2d language of \cite{Beem:2013sza}, the resulting states form a closed vertex algebra (VA) we shall refer to as the \lq\lq square root" VA since we only capture the states corresponding to $\lambda^1_{i,+}$ and derivatives
\begin{equation}
{\rm VOA}\ \xrightarrow{t\to0}\ {\rm VA}\cong\sqrt{\rm VOA}~.
\end{equation}

What happens in more general theories? Clearly all Lagrangian theories simply involve gauging collections of free fields (here we avoid cases without freely generated Coulomb branches) and therefore have a more restricted set of states, since the fields in \eqref{GCff} are no longer gauge invariant. Therefore, we have \cite{Gadde:2011uv}
\begin{equation}
\CI_{GC}(q,u)=\prod_r{\rm P.E.}\left({(u-q)u^{r-1}\over1-q}\right)~,
\end{equation}
where $r$ runs over the collection of $\bar\CE_r$ generators of the Coulomb branch. 

More generally, we can consider non-Lagrangian theories (or, at least, theories that do not have manifest $\CN=2$ Lagrangians). For example, it follows from the work of \cite{Bhargava:2022yik} that the minmal Argyres-Douglas theory (sometimes referred to as the $(A_1, A_2)$, $H_0$, or $\mathfrak{a}_0$ SCFT) has
\begin{equation}
\CI_{GC}(q,u)={\rm P.E.}\left({(u-q)u^{1/5}\over1-q}\right)~.
\end{equation}
In fact, using available $\CN=1$ Lagrangians, we have checked that the generalized Coulomb indices of the corresponding rank-one $H_0,H_1,H_2,D_4$ $\CN=2$ SCFTs all take the following universal form 
\be
\CI_{GC}(q,u)=
\frac{(q;q)_{\infty } \left(u^{2r-2};q\right){}_{\infty }}{2 \left({u^{r-1}};q\right)_{\infty } \left(u^{r};q\right){}_{\infty }}
\oint\frac{dz}{2\pi i z} 
\frac{  \left(z^{\pm 2};q\right){}_{\infty }}
{ \left( z^{\pm 2 } u^{r-1};q\right){}_{\infty }}~,
\ee
which can numerically be checked to be equal to\footnote{In the $H_0$ case, the expression below follows from the exact analysis of \cite{Bhargava:2022yik}.}
\begin{equation}
\CI_{GC}(q,u)={\rm P.E.}\left({(u-q)u^{r-1}\over1-q}\right)~,
\end{equation}
where $r$ is the scaling dimension of the generator of the Coulomb branch Chiral ring (this expression was also argued for via $S$-duality with $\CN=2$ Lagrangians in the case of the $E_{6,7,8}$ theories in \cite{Gadde:2011uv}). Furthermore, one can check that in these cases the index just captures the contributions from the following short multiplets
\be
\CI_{GC}(q,u)=1+ \sum_{n=1}^\infty\cI_{\bar\cE_{nr}}+\sum_{R,r,j,\bar j \atop
R+\bar j -j+1=0}\cI_{\bar \cC_{R,r(j,\bar j)}}~,
\ee
where the ${\bar \cC_{R,r(j,\bar j)}}$ multiplets subject to $R+\bar j -j+1=0 $ are the representations studied in  \cite{Bhargava:2022yik} (there an explicit and exact counting formula for this type of multiplet was given in the $H_0$ case).

Therefore, we are led to the following conjecture:

\bigskip
\noindent
{\bf Conjecture 4:} Any local unitary 4d $\CN=2$ SCFT with a freely generated Coulomb branch has 
\begin{equation}
\CI_{GC}(q,u)=\prod_r{\rm P.E.}\left({(u-q)u^{r-1}\over1-q}\right)~.
\end{equation}
In deriving consequences of this conjecture, {\it we will implicitly assume that any SCFT we discuss has a freely generated Coulomb branch} (apart from theories of free hypermultiplets, which do not have a Coulomb branch).

To that end, first note that, since the index is written in terms of plethystic exponentials of contributions from $\bar\CE_r$ multiplets (in the interacting case), and these multiplets cannot transform under flavor symmetry \cite{Buican:2013ica,Buican:2014qla}, the generalized Coulomb index is independent of flavor fugacities. In other words
\begin{equation}\label{GenGC}
\partial_a\CI_{GC}(q,u)=0~,
\end{equation}
where $a$ is any flavor fugacity.

\bigskip
\noindent
Since these indices are independent of flavor, it is natural to conjecture that for interacting theories:

\bigskip
\noindent
{\bf Conjecture 5:} Any multiplets in an irreducible local unitary interacting 4d $\CN=2$ SCFT contributing to the generalized Coulomb index are neutral under continuous flavor symmetries.\footnote{Otherwise their contributions would all need to cancel in the index. Here \lq\lq irreducible" means that the theory has a single sector / a single EM tensor multiplet.}

\bigskip
\noindent
Next, note that, in an interacting theory (again we consider theories without a decoupled free sector), $r>1$ in \eqref{GenGC}. As a result, we expect the corresponding reduced Macdonald index to be trivial (since $u^{r-1}\to0$). A similar statement holds for a free theory of hypermultiplets. Therefore, we have:

\bigskip
\noindent
{\bf Conjecture 6:} The reduced Macdonald index for a local unitary 4d $\CN=2$ SCFT, $\CT$, is non-trivial if and only if $\CT$ has (free) abelian vector multiplets.

\bigskip
\noindent
A related conjecture along the same lines is as follows:

\bigskip
\noindent
{\bf Conjecture 7:} There are non-trivial operators contributing to the reduced Macdonald index of a local unitary 4d $\CN=2$ SCFT, $\CT$, if and only if the theory has free vector multiplets (moreover, non-trivial contributions only come from gauge-invariant operators built out of free vector multiplets and derivatives). This statement also means that
\begin{equation}
\sqrt{\rm VOA}(\CT)\ne1\ \Leftrightarrow\ \CT\ {\rm has\ free\ vector\ multiplets}~,
\end{equation}
where \lq\lq1" denotes the trivial VA (moreover, the non-trivial states in $\sqrt{\rm VOA}$ only come from images of gauge-invariant operators built from free vector multiplets and derivatives).

\bigskip
\noindent
In light of \cite{Beem:2019tfp}, it is amusing to note that we are essentially saying that if $\sqrt{\rm VOA}$ is non-trivial, then it admits a free field realization.

\bigskip
Let us now connect the above conjectures to explicit constraints on representations of the SCA. To that end, we would like to know which multiplets contribute to the generalized Coulomb index. First, we have the contributions from $\bar\CE_r$ multiplets alluded to above
\begin{equation}
\CI_{\bar\CE_r}(q,u)={(u-q)u^{r-1}\over1-q}~.
\end{equation}

Next, consider the $\bar\CC$ multiplets. In the $t\to0$ limit we find
\begin{equation}
\CI_{\bar\CC_{R,r(j,\bar j)}}=(-1)^{2(j+\bar j)+1}{(u-1)q^{2j}(q-u)u^{\bar j-j+r}t^{\bar j-j+R+1}\over 1-q}+\cdots~,
\end{equation}
where we have dropped sub-leading terms in $t$. Clearly, the only operators that survive in the $t\to0$ limit are those satisfying $R=j-\bar j-1$
\begin{equation}
\CI_{\bar\CC_{j-\bar j-1,r(j,\bar j)}}(q,u)=(-1)^{2(j+\bar j)+1}{(u-1)q^{2j}(q-u)u^{\bar j-j+r}\over 1-q}~.
\end{equation}
Using the representation-theoretical identity $\bar\CB_{R,r(j,0)}\cong\bar\CC_{R-1/2,r-1/2(j,-1/2)}$ we have 
\begin{equation}
\CI_{\bar\CB_{R,r(j,0)}}=(-1)^{2j}{(u-1)q^{2j}(q-u)u^{-1-j+r}t^{-j+R}\over 1-q}+\cdots~,
\end{equation}
and so the contributing multiplets in the $t\to0$ limit are
\begin{equation}
\CI_{\bar\CB_{j,r(j,0)}}(q,u)=(-1)^{2j}{(u-1)q^{2j}(q-u)u^{-1-j+r}\over 1-q}~.
\end{equation}
In particular, this limit of the index captures $j=1/2$ multiplets known to exist under relatively minimal assumptions in higher-rank theories \cite{Banerjee:2023ddh} (these multiplets are indeed flavor neutral). But it also includes infinitely many generalizations.

Next, let us consider the $\hat\CC$ multiplets. We find
\begin{equation}
\CI_{\hat\CC_{R(j,\bar j)}}=(-1)^{2(j+\bar j)+1}{(u-1)q^{2j+1}t^{\bar j-j+R+1}\over 1-q}+\cdots~,
\end{equation}
and the surviving multiplets in the $t\to0$ limit are
\begin{equation}
\CI_{\hat\CC_{j-\bar j-1(j,\bar j)}}(q,u)=(-1)^{2(j+\bar j)+1}{(u-1)q^{2j+1}\over 1-q}~.
\end{equation}
For $j=\bar j+1$, these include the \lq\lq holomorphic" higher-spin multiplets discussed in \cite{Bhargava:2022yik} along with infinitely many more representations. Using the identity $\bar\CD_{R(j,0)}\cong\hat\CC_{R-1/2(j,-1/2)}$ we also have
\begin{equation}
\CI_{\bar\CD_{R(j,0)}}=(-1)^{2j}{(u-1)q^{2j+1}t^{R-j}\over 1-q}+\cdots~,
\end{equation}
along with the surviving multiplets contributing to the generalized Coulomb limit
\begin{equation}
\CI_{\bar\CD_{j(j,0)}}(q,u)=(-1)^{2j}{(u-1)q^{2j+1}\over 1-q}~.
\end{equation}
Using the identity $\hat\CB_R\cong\hat\CC_{R-1(-1/2,-1/2)}$, we see that, as expected from our free-field discussion, $\hat\CB_R$ multiplets do not contribute to the generalized Coulomb limit.

From these results, we see that our above conjectures imply the following additional conjecture

\bigskip
\noindent
{\bf Conjecture 8:} The following multiplets (and their conjugates) are neutral under any continuous flavor symmetry in an irreducible local unitary interacting 4d $\CN=2$ SCFT
\begin{equation}\label{flavneut}
\bar\CC_{j-\bar j-1,r(j,\bar j)}\oplus\bar\CB_{j,r(j,0)}~.
\end{equation}

\bigskip
\noindent
It is crucial that the above multiplets saturate our bound, $j=\bar j+R+1$. Indeed, otherwise we could consider taking primaries saturating this bound and taking the normal-ordered product with the primary of a Noether current mulitplet for a non-abelian symmetry in a decoupled sector, $\hat\CB_1$. This procedure would then produce multiplets transforming under flavor, which would be a contradiction. Note that known examples of multiplets that transform under flavor symmetry are not of the above type. For example, $\bar\CD_{1/2(0,0)}$ contains extra supercurrents and therefore transforms under a $U(1)$ flavor symmetry (in the case of $\CN=3$ and $SU(2)$ in the case of $\CN=4$), but this representation is not in \eqref{flavneut}. Moreover, neither is the $U(1)$-charged $\bar\CB_{1,2(0,0)}\in\bar\CD_{1/2(0,0)}\times\bar\CD_{1/2(0,0)}$ multiplet required by locality and unitarity \cite{Banerjee:2023ddh}.

\bigskip
Next, let us consider taking $u\to0$ and going to the reduced Macdonald index corresponding to the $\sqrt{\rm VOA}$ vertex algebra. The surviving multiplets and index contributions are
\begin{equation}
\CI_{\hat\CC_{j-\bar j-1(j,\bar j)}}(q)=(-1)^{2(j+\bar j)}{q^{2j+1}\over 1-q}~,\ \ \ \CI_{\bar\CD_{j(j,0)}}(q)=(-1)^{2j+1}{q^{2j+1}\over 1-q}~.
\end{equation}
From our conjecture that such multiplets are present only in theories with free vectors, we obtain the following conjecture:

\bigskip
\noindent
{\bf Conjecture 9:} The following representations (and their conjugates) are present if and only if a 4d $\CN=2$ SCFT consists of free vector multiplets (or contains such a sector)
\begin{equation}\label{freeconj}
\hat\CC_{j-\bar j-1(j,\bar j)}\oplus \bar\CD_{j(j,0)}~.
\end{equation}

\bigskip
\noindent
Typically, the $j=\bar j+1$ $\hat\CC$ case is associated with higher-spin symmetry and hence free vector theories. On the other hand, here we see that all $\hat\CC$ multiplets in \eqref{freeconj} are diagnostics of free theories. We conclude by noting a peculiar feature of the above representations. Let us denote $j_{\rm max}$ as the maximum spin in the $\bar\CD_{j(j,0)}$ sector. For simplicity, suppose our theory consists of $n_v$ free abelian vector multiplets and $n_H$ free hypermultiplets. It is easy to check that $j_{\rm max}=(n_v-1)/2$. In other words, we have
\begin{equation}
j_{\rm max}= 2(2a-c)-1/2~,
\end{equation}
where $a$ and $c$ are the central charges of our free theory. It is interesting to note that, albeit in a limited sense, $\sqrt{\rm VOA}$ \lq\lq knows" about this combination of 4d central charges.

\newsec{5d $\CN=1$}
Let us now consider 5d $\CN=1$.\footnote{Recall that this is the only amount of SUSY compatible with superconformal invariance \cite{Nahm:1977tg}.} In this number of spacetime dimensions, IR Lagrangians are quite powerful in the sense that they can often be used to derive indices of UV SCFTs \cite{Kim:2012gu} (although see \cite{Kim:2023qwh} for some limitations). On the other hand, these Lagrangians are typically not conformal due to the fact that, as in 3d, free vectors are not conformal. However, unlike in 3d, there are no twisted hypermultiplets, and so the only free representation is the hypermultiplet. Therefore, we expect free fields to be less powerful in 5d.

To see this heuristic reasoning is indeed correct, note that the hypermultiplet primary has $R=1$ and $\Delta=3/2$, while the fermion has $\Delta=2$ and $R=0$ (here we follow the general logic of section 3 of \cite{Buican:2016hpb}). They sit in a $C_1[0,0]_{3/2}^{(1)}$ multiplet in the notation of \cite{Cordova:2016emh} (or a $\bar\CD[0,0;1]$ multiplet in the notation of \cite{Buican:2016hpb}). 

Now, consider the $A_4[0,0]^{(0)}_{4}$ representation (or a $\CA[0,0;0]$ multiplet in the language of \cite{Buican:2016hpb}).\footnote{This short representation is interesting because it houses the only short primary that can potentially get a vev in flowing to a Coulomb branch.} This multiplet has a scalar primary of dimension four. To construct it from free fields, we must have either zero or two bosons. Suppose we try with two. Then, the remaining scaling dimension must be made up by a derivative, but the spin of the primary will not be zero. Therefore, suppose we have zero bosons. We must have an even number of fermions: zero or two. Clearly we cannot have zero, and so we have two fermions. Then, we have an operator schematically of the form $\psi_1\psi_2$, but this cannot be a primary. 

Do such primaries exist in interacting theories? In \cite{Buican:2016hpb}, the answer was shown to be yes by analyzing the index of the $E_2$ SCFT found in \cite{Kim:2012gu} (this SCFT was originally discovered in \cite{Seiberg:1996bd}). This discussion is therefore compatible with the conjectures and ideas presented in this paper since we expect higher-dimensional theories to be \lq\lq less" Lagrangian, and there is no continuous deformation of $\CA_4[0,0]_4^{(0)}$ to an irrep realized in a free theory.
 
 \newsec{Conclusion}
 In this paper we have studied the question of when it is possible to realize all unitary representations of SCAs with eight Poincar\'e supercharges compatible with locality via free fields in $d=2\to5$. As we have seen, the answers (and conjectures) reproduce certain heuristic QFT structures we know from experience (e.g., the increasing power of Lagrangians as we lower the spacetime dimension). Moreover, our discussion touches on the important principles of unitarity and locality. By its nature, our work leaves various open questions for future exploration. For example:
 \begin{itemize}
\item Use the ANEC or some similar relation to prove our conjectures in 4d (or, alternatively, find counterexamples).\footnote{We are currently exploring this direction for the $\bar\CB$ conjecture \cite{toApp}.}
\item Make more concrete the connection of our results with the topology of the space of QFTs, $\CT_{\rm QFT}$. Since our question is representation theoretical in nature, it should be possible to phrase in terms of the superconformal index and, perhaps, some more powerful index that more tightly constrains the topology of $\CT_{\rm QFT}$ (or at least the subspace of theories with eight Poincar\'e supercharges). Perhaps in this way it will be possible to make contact with ideas in \cite{Vafa:1988ue,Gukov:2016tnp}.
\item Studying the same question in 6d $(1,0)$ will, in some sense, complete our study (since this is the maximal dimension for SCFTs).
\item Understand how our discussion interfaces with category theory (e.g., see the recent work \cite{Zeev:2022cnv}).
\item If we allow for differing amounts of supersymmetry, can we shed new light on supersymmetry enhancement?
\item To what extent can free fields in $d>2$ also reproduce OPE coefficients of interacting theories? There are some results in this direction (at least for vanishing OPE coefficients) in \cite{Bhargava:2022cuf,Bhargava:2022yik,Bhargava:2023zbd}.
\item Spin played an important role in all of our bounds. What happens if we relax rotational invariance and ask similar questions? Continuum theories without continuous rotational symmetry have been much-studied in recent years (e.g., see \cite{Seiberg:2020bhn} and references therein).
\item If a particular SCA irrep does not have a realization in terms of free fields (e.g., as in the case of 3d $\CN=4$ long multiplets with $\Delta\not\in\mathbb{Z}/2$), is there a sense in which there is a \lq\lq closest" free field irrep? Can this notion of closeness be related to geodesics in the space of theories? Can it be related to a property of conformal interfaces?
\item We saw that the maximal spin for the Hall-Littlewood sub sector of 4d irreps related to $\sqrt{\rm VOA}$ knew about a combination of central charges in certain theories. Are there other irreps whose maximal spin quantum number characterise combinations of central charges in more general theories? Can this lead to a free-field embedding theorem along the lines conjectured in \cite{Douglas:2010ic} for 2d theories?\footnote{We thank A.~Banerjee for emphasizing the significance of the conjecture in \cite{Douglas:2010ic} to us.}
 \end{itemize}
 We hope to return to some of these questions soon.
 
 \ack{We are grateful to A.~Banerjee, C.~Bhargava, J.~Distler, A.~Manenti, and S.~Razamat for comments and discussions. M.~B. and H.~J. were supported by the grant “Amplitudes, Strings and Duality” from STFC.  M.~B. was also supported by the grant “Relations, Transformations, and Emergence in Quantum Field Theory” from the Royal Society. M.~B. thanks SISSA and Nordita for hospitality during parts of this project. No new data were generated or analysed during this study.}
 
\newpage

\bibliography{chetdocbib}

\end{document}